\documentclass[conference]{IEEEtran}
\usepackage{amsmath,graphicx,epstopdf}
\usepackage{multicol}
\usepackage{url}
\usepackage{cite}
\usepackage{multirow}

\ifCLASSINFOpdf
\else
\fi
\begin{document}

%
\title{Raw Multi-Channel Audio Source Separation using Multi-Resolution Convolutional Auto-Encoders}

\author{\IEEEauthorblockN{Emad M. Grais, Dominic Ward, and Mark D. Plumbley}
\IEEEauthorblockA{Centre for Vision, Speech and Signal Processing,\\University of Surrey, Guildford, UK.\\
Email: {grais, dominic.ward, m.plumbley}@surrey.ac.uk}
}


%


\maketitle

\begin{abstract}
Supervised multi-channel audio source separation requires extracting useful spectral, temporal, and spatial features from the mixed signals. The success of many existing systems is therefore largely dependent on the choice of features used for training. In this work, we introduce a novel multi-channel, multi-resolution convolutional auto-encoder neural network that works on raw time-domain signals to determine appropriate multi-resolution features for separating the singing-voice from stereo music. Our experimental results show that the proposed method can achieve multi-channel audio source separation without the need for hand-crafted features or any pre- or post-processing.

\end{abstract}


%
\IEEEpeerreviewmaketitle

\section{Introduction}
\label{sec:intro}
In supervised multi-channel audio source separation (MCASS), extracting suitable
spectral, temporal, and spatial features is usually the first step toward
tackling the problem
\cite{arie:16:masswdnn,Stefan:17:imssbdnntdanb,Vincent:06:mssutfsp}. The
spectro-temporal information is considered imperative for discriminating between
the component sources, while spatial information can be harnessed to achieve
further separation \cite{Yu:16:lbssssptfmdnn,Zermini:17:blpsdnnsns}. The
spectro-temporal information is typically extracted using the short-time Fourier
transform (STFT), where there is a trade-off between frequency and time
resolutions \cite{Griffin:84:semstft}. Computing the STFT to obtain features with
high resolution in frequency leads to features with low resolution in time, and
vise versa \cite{Griffin:84:semstft}. Most audio processing approaches prefer an 
auditory motivated frequency scale such as Mel, Bark, or Log scaling rather than a linear frequency scale \cite{gao:13:uscsnsgfinmt,Choi:17:tdlmir}. However, it is usually not easy to reconstruct the time-domain signals from those type of features. Another common pre-processing step is to take the logarithm of the spectrograms. Despite this, many source separation techniques focus on estimating the magnitude spectra, using the phase of the mixture to reconstruct the time-domain source signals \cite{Emad:16:scassdnne,Zermini:17:blpsdnnsns}. Unfortunately, omitting phase estimation for the sources usually results in poor perceptual separation quality \cite{Dubey:17:dpmmss,Krawczyk:14:stftphase}. Spatial information can be extracted for example from the magnitude and phase differences of the STFT of different spatial channels \cite{Yu:16:lbssssptfmdnn,Zermini:17:blpsdnnsns}, or by estimating a spatial covariance matrix \cite{arie:16:masswdnn,Stefan:17:imssbdnntdanb}. All the aforementioned features are hand-crafted features and most of the time we can not have features that are good in representing all the spectral, temporal, and spatial characteristics of different audio sources. There is usually a trade-off between these features.

Instead of humans deciding which features to extract from the audio signals, recently, different deep neural networks (DNNs) have been used to process the time-domain audio signal directly to automatically extract the suitable features for each type of audio signals \cite{Sainath:15:lsferw,Sainath:15:mspdnnasr,Dieleman:14:eelma,Venkataramani:17:afeeess,fu:17:eewuedemofcnn,Yedid:15:samrmw}. In those papers, convolutional layers in the DNNs were capable of extracting useful features from the raw waveforms of the input signal. Each convolutional layer in \cite{Sainath:15:lsferw,Sainath:15:mspdnnasr,Dieleman:14:eelma,Venkataramani:17:afeeess,fu:17:eewuedemofcnn,Yedid:15:samrmw} has filters with the same size, which extract features with a certain time resolution.    

In this paper, we propose a novel multi-channel Multi-Resolution Convolutional Auto-Encoder (MRCAE) neural networks for MCASS. Each layer in MRCAE is composed of sets of filters, where the filters in one set have the same size which is different to the sizes of the filters in the other sets. The large filters extract global information from the input signal while small filters extract the local details from the input signal. The features that capture both global and local (multi-resolution) details can help discriminating between different audio sources, which is an essential issue for source separation. The inputs and outputs of the MRCAE are the mixtures and the estimated target sources respectively in the time-domain. The proposed MRCAE is also multi-channel which captures the information in the different channels of the input signals. We do not perform any pre-processing or post-processing operations on the audio signals. 
%

This paper is organized as follows. In Section \ref{sec_mrcae}, the proposed MRCAE neural network is presented. In Section \ref{mrcae_ss}, we show how the proposed MRCAE is used for source separation. The remaining sections present the experiments and conclusion of our work.

\section{Multi-resolution convolutional auto-encoder neural networks}
\label{sec_mrcae}
\begin{figure}[t!]
 \includegraphics[width=1\linewidth,height=4.1cm]{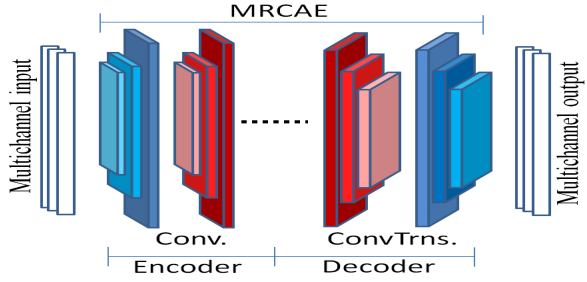}
\caption{\label{mrcae}{Overview of the structure of a multi-channel multi-resolution convolutional auto-encoder (MRCAE). ``Conv'' denotes convolutional layers and ``ConvTrns'' denotes transpose convolutional layers. Each layer consists of different sets of filters with different sizes.
}}
\end{figure}

The proposed multi-resolution convolutional auto-encoder (MRCAE) neural network is a fully convolutional denoising auto-encoder neural network as in \cite{grais:17:scasscda}, but with each layer consisting of a different set of filters. The MRCAE has two main parts, the encoder and decoder parts. The encoder is used to extract multi-resolution features from the input mixtures and the decoder uses these features to estimate the sources. The encoder and decoder consist of many convolutional and transpose convolutional layers \cite{Dumoulin:16:gcadl} respectively as shown in Fig \ref{mrcae}. Each layer in MRCAE consists of different sets of filters, where the filters in one set have the same size and the filters in different sets have different sizes.

Considering the concept of calculating the STFT of an audio signal, if the STFT window is large, the STFT features capture the frequency components of the signal in high resolution and the temporal characteristics in low resolution \cite{Griffin:84:semstft} and vice versa. STFT can not produce features in high resolution in both time and frequency. 

\begin{figure}[t!]
\begin{center}
 \includegraphics[width=1\linewidth,height=6cm]{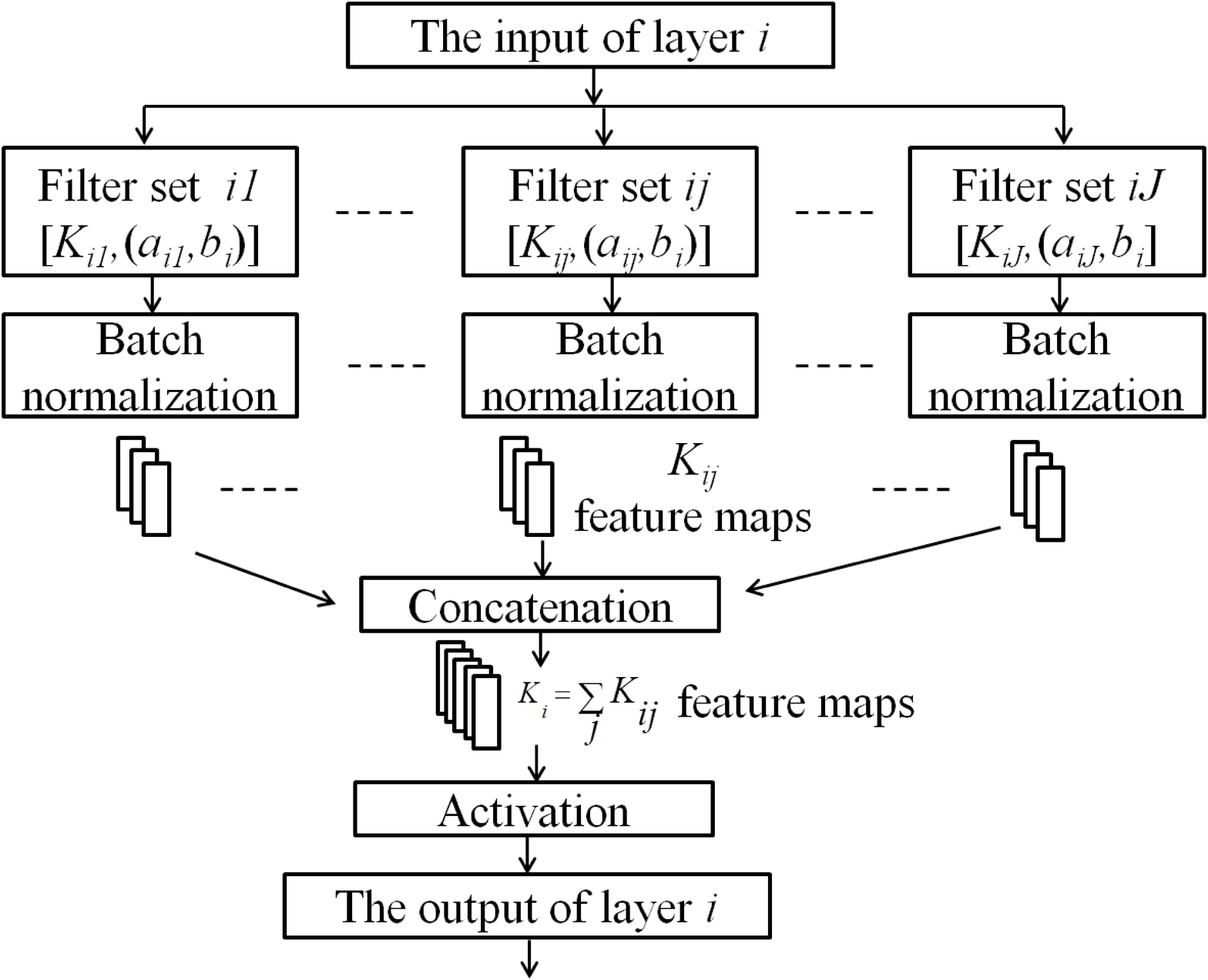}
\caption{\label{fig:layer_mrfcnn}{Overview of the proposed structure of each layer of the MRCAE. Where $K_{ij}$ denotes the number of filters with size $a_{ij} \times b_{i}$ in set $j$ in layer $i$, $a_{ij}$ is the length of the filters in the time direction, and $b_{i}$ is the size of the filters that equals to the number of channels in the input. ``Activation'' denotes the activation function. 
}}
\end{center}
\end{figure}

To build a system that is automatically capable of extracting suitable features from the input raw data (time-domain signal) in a suitable time and frequency resolution according to each source in the input mixtures, we propose to use MRCAE, where each layer consists of different sets of filters with different sizes as shown in Fig. \ref{fig:layer_mrfcnn}. This figure shows that at each layer $i$ there are $J$ sets of filters. Each filter set $j$ in layer $i$ has $K_{ij}$ filters with the same size $a_{ij}\times b_i$, where $a_{ij}$ is the filter length and $b_i$ is the number of channels that the input data to layer $i$ has. In each layer $i$, the value of $a_{ij}$ in set $j$ is different than the value $a_{ij'}$ in set $j'$, but $b_i$ is the same for all sets in the same layer $i$, because all sets have the same number of channels of the input data to the same layer. Each set $j$ of filters at layer $i$ generates $K_{ij}$ feature maps in a certain resolution and each layer $i$ generates $K_i=\sum_j^JK_{ij}$ feature maps in different resolutions. The $K_i$ is the number of channels for the input data of the next layer. 

The long filters with large $a_{ij}$ are good in capturing the global information of the processed signals and the short filters with small $a_{ij}$ can capture the local details. We might think of using long filters as calculating STFT over long window, and the short filters as calculating STFT over short window. This means using long and short filters together in the same layer produces features with different time-frequency resolutions. This can be very useful for many audio signal processing applications. In MCASS, there are different audio sources in the mixtures and useful information can be extracted for different sources using different time-frequency resolutions that is suitable for different sources. Since the input signal is multi-channel time-domain signal, each filter in the first layer is a multi-dimensional filter to be able to run over the multi-channel input signals. 
%
\section{MRCAE for multi-channel audio source separation}
\label{mrcae_ss}
\label{overall}
Suppose we have $C$ mixtures each with $L$ sources as $y(t,c) = \sum_{l=1}^L s_l(t,c)$, $\forall$ $c \in C$, where $C$ is the number of channels and $t$ denotes time. The aim of MCASS is to estimate the sources $s_l(t,c), \ \forall{l,c}$, from the mixed signals $y(t,c)  \ \forall c$. In the stereo case, $C=2$. We work here on the time-domain input and output signals.

In this work, we propose to use a single MRCAE to separate all the target sources from the input mixtures. The inputs for the MRCAE is multi-channel (two channels for the stereo case) segments of the input mixture signal. Each segment has length $N$ of time-domain samples. The corresponding output segment for each target source is also multi-channels with length $N$ samples. The total number of filters in the output layer of the MRCAE should be equal to the number of target sources multiplied by the number of channels for each source. This way we guarantee that the output layer generates feature maps equal to the number of target sources, where each source has its multiple channel components. For example, in the stereo case, if we wish to separate four sources, the number of filters in the output layer should be eight filters.    
\subsection{Training the MRCAE for source separation}
\label{sec:train}
Let us assume we have training data for the mixed signals and their corresponding target sources. Let $y(t,c)$ be the mixed input signal for channel $c$ and $s_l(t,c)$ be the target source $l$ for channel $c$. The MRCAE is trained to minimize the following cost function:
\begin{equation}
\label{cost_mask}
D =\sum_{t,c,l}\left| z_l(t,c) - s_l(t,c) \right|
\end{equation}
where $z_l(t,c)$ is the actual output of the last layer of the MRCAE for source $l$ and channel $c$, $s(t,c)$ is the reference target output signal for source $l$ and channel $c$. The input of the MRCAE is the mixed signals $y(t,c),  \ \forall c$. 
\subsection{Testing the MRCAE for source separation}
\label{sec:test} 
The multi-channel mixture is passed through the trained MRCAE. The output of each filter in the last layer is considered to be the time-domain estimate of one of the channels $c$ of one of the sources $l$.
%
\section{Experiments}
\label{sec:exp}
We applied our proposed MRCAE approach to separate the singing-voice/vocal sources from a group of songs from the SiSEC-2016-MUS-task dataset \cite{Liutkus:17:ssec}. The dataset has 100 stereo songs with different genres and instrumentations. Each song is a mixture of vocals, bass, drums, and other musical instruments. The first 50 songs in the dataset were used as training and validation datasets, and the last 46 songs were used for testing. The data were sampled at 44.1kHz. 

The quality of the separated vocals was measured using four metrics of the BSS-Eval toolkit \cite{vincent:06:pmi}: source to distortion ratio (SDR), source image to spatial distortion ratios (ISR), source to interference ratio (SIR), and source to artifacts ratio (SAR). ISR is related to the spatial distortion, SIR indicates the remaining interference between the sources after separation, and SAR indicates the artifacts in the estimated sources. SDR measures the overall distortion (spatial, interference, and artifacts) of the separated sources, and is usually considered the overall performance evaluation for any source separation approach \cite{vincent:06:pmi}. Achieving high SDR, ISR, SIR, and SAR indicates good separation performance.

In the training stage of the MRCAE, the time-domain samples of the 50 signals for the input mixtures from the training set were normalized to have zero mean and unit variance. The normalized input mixtures and their corresponding target vocal source were then divided into segments of length 1025 samples in each segment. The segments of the input mixtures and the target vocal signals were used to train the MRCAE. 

In the test phase, the input signals of each song were divided into 1025 samples with hop size 16 and passed through the trained MRCAE. The outputs of the MRCAE were used with simple shift and add procedures to reconstruct the estimate for the time-domain signal for the target vocal source. It is worth mentioning that we did not perform any pre- or post-processing on the input or output data other than normalizing the input signals to have zero mean and unit variance.

\subsection{MRCAE structure}
\label{sec:parameters}
The MRCAE consists of two convolutional layers in the encoder part, two transpose convolutional \cite{Dumoulin:16:gcadl} layers in the decoder part, and one output layer as shown in Table \ref{table:cdae}. Table \ref{table:cdae} also shows the number of filter sets, the number of filters in each set, and the length of the filters in each set. The choices for filter length as an analogy for calculating the STFT with different window sizes as 5, 50, 256, 512, and 1025. 
The short filters capture the local details in high resolution in time while the long filters capture the global information (maybe seen as features with high frequency resolution) of the input signals. Since we separate one source (vocal) with two channels, the output layer of the MRCAE is a transpose convolutional layer with two filters, where each filter generates a feature map corresponding to the estimate of one of the channels of the estimated vocal. Batch normalization was used after each set of filters as shown in Fig. \ref{fig:layer_mrfcnn}. The activation function for all layers is exponential linear unit (ELU) function that allows positive and negative values in its output, which has been shown to speed up the learning in deep neural networks \cite{Clevert:15:elu}. The length of the input and output segments for the MRCAE was 1025 time-domain samples. 
%
%
\begin{table}[t!]
\begin{center}
\scalebox{0.82}
{
\begin{tabular}{||p{.55cm}|p{.55cm} p{1.8cm}|p{.55cm} p{2.3cm}|p{2.2cm}||}
 \hline
 \multicolumn{6}{|c|}{MRCAE model summary. The input/output data with size 1025 samples} \\
 \hline
Layer & \multicolumn{2}{|c|}{Encoder} &\multicolumn{2}{|c|}{Decoder} &Output\\
  \hline
\multirow{5}{*}{1} & set 1 & Conv[20,(5)]    & set 1 & ConvTrns[50,(5)]   & \multirow{5}{*}{ConvTrns[2,(1025)]} \\
                   & set 2 & Conv[20,(50)]   & set 2 & ConvTrns[25,(50)] &\\ 
                   & set 3 & Conv[20,(256)]  & set 3 & ConvTrns[20,(256)] & \\
                   & set 4 & Conv[20,(512)]  & set 4 & ConvTrns[20,(512)] & \\
                   & set 5 & Conv[20,(1025)] & set 5 & ConvTrns[20,(1025)] & \\
\hline 
\multirow{5}{*}{2} & set 1 & Conv[50,(5)]    & set 1 & ConvTrns[20,(5)]   & \multirow{5}{*}{ } \\
                   & set 2 & Conv[25,(50)]   & set 2 & ConvTrns[20,(50)] &\\ 
                   & set 3 & Conv[20,(256)]  & set 3 & ConvTrns[20,(256)] &\\
                   & set 4 & Conv[20,(512)]  & set 4 & ConvTrns[20,(512)] &\\
                   & set 5 & Conv[20,(1025)] & set 5 & ConvTrns[20,(1025)] &\\
                 
 \hline
\end{tabular}
}
\caption{The detail information about the number and sizes of the filters in each layer in the MRCAE. For example ``Conv[20,(5)]'' denotes convolutional layer with 20 filters and the length of each filter is 5. ``ConvTrns'' denotes transpose convolutional layer.}
\label{table:cdae} 
\end{center}
\end{table}

The parameters for the MRCAE were initialized randomly. The MRCAE was trained using backpropagation with gradient descent optimization using Adam \cite{adam:14:amso} with parameters $\beta_1=0.9$, $\beta_2=0.999$, $\epsilon=1e-08$, batch size 100, and a learning rate of $0.0001$, which was reduced by a factor of 10 when the values of the cost function ceased to decrease on the validation set for 3 consecutive epochs. The maximum number of epochs was 20. We implemented our proposed algorithm using Keras with Tensorflow backend \cite{chollet2015keras2}.
\begin{figure*}[h!]
 \centering
 \includegraphics{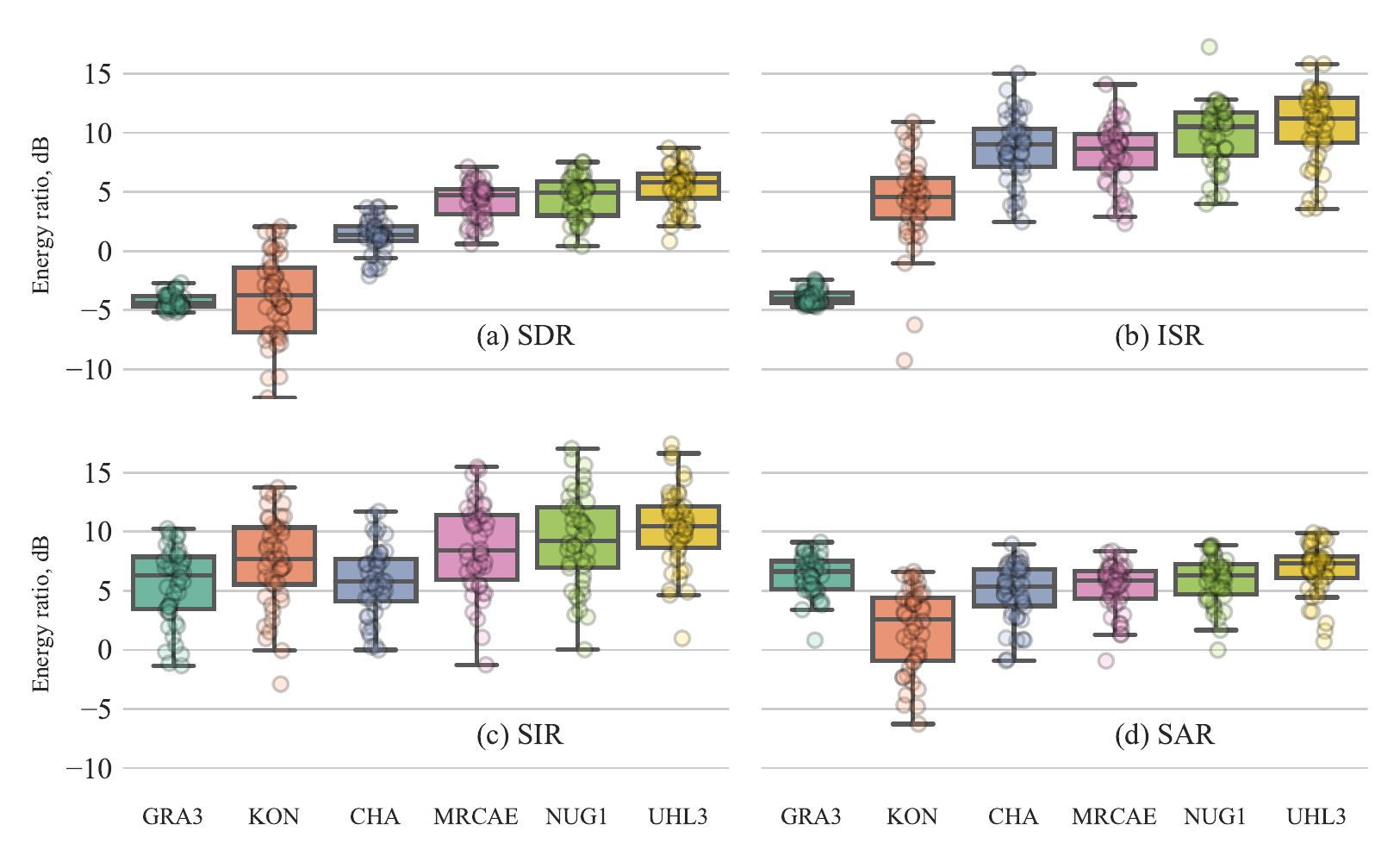}
\caption{\label{fig:sdr}{Boxplots (with individual data points overlaid) of the SDR (a), ISR (b), SIR (c) and SAR (d) BSS-Eval performance measures for our proposed MRCAE and five singing-voice separation systems applied to the SiSEC-2016-MUS test set.}}
\end{figure*}
\subsection{Comparison with related works}
\label{sec:results}
We compared the performance of the proposed MRCAE approach for MCASS with five different deep neural networks (DNNs) based approaches from the submitted results to the SISEC-2016-MUS challenge \cite{Liutkus:17:ssec}. Two of those approaches are the best submitted results in this challenge, known as UHL3 and NUG1 in \cite{Liutkus:17:ssec}, and the three other approaches are known as CHA, KON, and GRA3 in \cite{Liutkus:17:ssec}. UHL3 combined different deep feed forward neural networks (FFN) and deep bidirectional long short-term memory (BLSTM) neural networks, with data augmentation from different data set \cite{Stefan:17:imssbdnntdanb}. In UHL3 the spectrogram of the linear combination of the outputs of the models was used to compute spatial covariance matrices to separate the sources from the input mixtures in the STFT domain. The second best approach in the SISEC-2016-MUS challenge was NUG1, which used deep FFN to find spectrogram estimates for the sources then these estimates were used to compute spatial covariance matrices that were then used to separate the sources in the STFT domain \cite{arie:16:masswdnn}. NUG1 used the expectation maximization (EM) algorithm to iterate between using the FFN to find spectrogram estimates and updating the spatial covariance matrices to improve the separation quality of the estimated sources. UHL3 and NUG1 stacked numbers of neighbouring frames of the spectrograms of the input mixtures and used principle component analysis (PCA) to reduce the dimensionality of the stacked spectral frames. CHA \cite{chandna:17:massudcnn} and KON used deep convolutional neural networks and deep recurrent neural networks respectively to extract the spectrogram of each source from the spectrogram of the average of the two channel input mixtures. GRA3 stacked the magnitude spectrograms of the two channels and used deep FFN to estimate the magnitude spectrograms of the two channels of each source \cite{Emad:16:scassdnne}. 
%
\subsection{Results}
\label{sec:results}
Fig. \ref{fig:sdr} shows boxplots of the SDR (a), ISR (b), SIR (c) and SAR (d) measures, of the proposed MRCAE method and the aforementioned five other DNN methods from the SISEC-2016-MUS challenge. Considering the SDR as the overall quality measurement, we can see that the proposed MRCAE method that works by just sending the mixed signals in the time-domain into the trained MRCAE to estimate the time-domain vocal signals works better than CHA, KON, and GRA3 that used STFT and different DNNs to estimate the sources. The performance of MRCAE in SDR, SIR, and SAR is not too far from UHL3 and NUG1 methods. The main advantage of our proposed approach over UHL3 and NUG1 is dealing with the raw data without any pre- or post-processing of the input and output signals. The works of UHL3 and NUG1 require many pre- and post processing such as: computing STFT and dealing with complex numbers, stacking numbers of neighbouring spectral frames, using PCA for dimensionality reduction, computing spatial covariance matrices, combining different DNN outputs, data augmentations, and iterative EM algorithm. The results in Fig. \ref{fig:sdr} shows that our proposed approach of using MRCAE for MCASS is very promising. In our future work, we hope by having better choices for the MRCAE parameters and better choice for the cost function than the shown one in Eq. \ref{cost_mask}, we can achieve better results than the shown ones in Fig. \ref{fig:sdr}. 

Table \ref{all_sisec2016} shows the across-song medians of the BSS-Eval measures for the proposed MRCAE and most of the submitted approaches to SiSEC-2016-MUS challenge \cite{Liutkus:17:ssec}. The order of the methods in Table \ref{all_sisec2016} is based on the SDR values. DUR \cite{Durrieu:12:ammrpemass}, KAM \cite{Liutkus:15:saslkam}, OZE \cite{Ozerov:12:gffhpias}, RAF3 \cite{Rafii:13:rpet}, JEO2 \cite{Jeong:17:svsrpca}, and HUA \cite{Huang:12:svsmrrp} are blind source separation approaches. STO1 \cite{Stoter:16:cfmuss} is supervised source separation approach based on feed-forward DNN architecture using patched overlapped STFT frames on input and output. According to the median SDR values, our proposed MRCAE outperforms most of the other approaches except UHL3 and NUG1. The difference in median SDR between MRCAE and UHL3 is -1dB and between MRCAE and NUG1 is -0.2dB. Audio examples of source separation using MRCAE are available online\footnote{\footnotesize{\url{https://cvssp.github.io/maruss-website/publications/Grais_2018.html}}}.
\begin{table}[t!]
\centering 
\begin{tabular}{c c c c c} 
\hline\hline 
Method & SDR & ISR & SIR & SAR \\ [0.5ex] 
\hline 
UHL3 & 5.79 &  11.23& 10.46 & 7.32\\ 
NUG1 & 4.91 &  10.52&  9.21& 6.30\\
\textbf{MRCAE}& 4.71 &  8.67&  8.43& 5.89\\
STO1 & 4.23 &  8.07&  8.44& 5.42\\
JEO2 & 4.20 &  8.76&  7.01& 5.91\\
KAM1 & 2.11 &  5.98 &  9.85& 1.09\\
RAF3 & 1.92 &  8.60&  1.42& 6.46\\
OZE & 1.85  &  5.46&  3.75& 2.18\\
DUR  & 1.36 &  1.57&  5.14& 2.86\\
CHA  & 1.34 &  9.05&  5.77& 5.38\\
KON  & -3.75&  4.56&  7.70& 2.59\\
HUA  & -4.14&  15.05&  -2.43& 7.99\\
GRA3 & -4.43&  -4.05&  6.31& 6.62\\ [1ex] 
\hline 
\end{tabular}
\caption{The median values for the BSS-Eval measures for our proposed MRCAE and most submitted systems to the SiSEC-2016-MUS test set.} 
\label{all_sisec2016} 
\end{table}
%
%
\section{Conclusion}
In this paper, we proposed a new multi-channel audio source separation method based on separating the waveform directly in the time-domain without extracting any hand-crafted features. We introduced a novel multi-resolution convolutional auto-encoder neural network to separate the stereo waveforms of the target sources from the input stereo mixed signals. Our experimental results show that the proposed approach is very promising. In future work we will investigate combining the multi-resolution concept with generative adversarial neural networks (GANs) for waveform audio source separation.    
\section*{Acknowledgment}
This work is supported by grant EP/L027119/2 from the UK Engineering and Physical Sciences Research Council (EPSRC).



\bibliographystyle{IEEEtran}
%
\bibliographystyle{IEEEtran.bst}
\bibliography{refs}

\end{document}